\RequirePackage{lineno} 
\documentclass{nature}
\usepackage{xcolor}
\usepackage{caption}
\usepackage{multirow}
\usepackage{amsmath}
\usepackage{CJK}
\usepackage{bm}
\usepackage{braket}
\usepackage{float}
\usepackage{amssymb}
\usepackage{color}
\usepackage{array}
\usepackage{ulem}
\usepackage{siunitx}
\usepackage{booktabs}
\usepackage{caption}
\usepackage{placeins}

\DeclareSIUnit\angstrom{\text{Å}}

\title{Moir\'e enhanced flat band in rhombohedral graphene}

\author{Hongyun Zhang$^{1,2,\dagger}$, Jinxi Lu$^{1,\dagger}$, Kai Liu$^{3,\dagger}$, Yijie Wang$^{4}$, Fei Wang$^1$, Size Wu$^{3}$, Wanying Chen$^1$, Xuanxi Cai$^{1}$, Kenji Watanabe$^5$, Takashi Taniguchi$^6$, Jose Avila$^{7}$, Pavel Dudin$^{7}$, Matthew D. Watson$^{8}$, Alex Louat$^{8}$, Takafumi Sato$^{2,9}$, Pu Yu$^{1,10}$, Wenhui Duan$^{1,10,11}$, Zhida Song$^{4}$, Guorui Chen$^{3,*}$ \& Shuyun Zhou$^{1,10,*}$}

\usepackage{graphicx}
\makeatletter
\let\saved@includegraphics\includegraphics
\AtBeginDocument{\let\includegraphics\saved@includegraphics}

\makeatother

\begin{document}
\maketitle

\begin{affiliations}
    \item State Key Laboratory of Low-Dimensional Quantum Physics and Department of Physics, Tsinghua University, Beijing 100084, P. R. China
    \item Advanced Institute for Materials Research (WPI-AIMR), Tohoku University, Sendai 980-8577, Japan
    \item Key Laboratory of Artificial Structures and Quantum Control (Ministry of Education), School of Physics and Astronomy and Tsung-Dao Lee Institute, Shanghai Jiao Tong University, Shanghai, P. R. China
    \item International Center for Quantum Materials, School of Physics, Peking University, Beijing, P. R. China
    \item Research Center for Electronic and Optical Materials, National Institute for Materials Science, 1-1 Namiki, Tsukuba 305-0044, Japan
    \item Research Center for Materials Nanoarchitectonics, National Institute for Materials Science,  1-1 Namiki, Tsukuba 305-0044, Japan
    \item Synchrotron SOLEIL, L’Orme des Merisiers, Saint Aubin-BP 48, 91192 Gif sur Yvette Cedex, France
	\item Diamond Light Source Ltd, Harwell Science and Innovation Campus, Didcot, OX11 0DE, UK
	\item Department of Physics, Graduate School of Science, Tohoku University, Sendai 980-8577, Japan
	\item Frontier Science Center for Quantum Information, Beijing 100084, P. R. China
    \item Institute for Advanced Study, Tsinghua University, Beijing 100084, P. R. China\\
    $\dagger$ These authors contributed equally to this work.\\
    *Correspondence should be sent to chenguorui@sjtu.edu.cn \& syzhou@mail.tsinghua.edu.cn
\end{affiliations}

\begin{abstract}

The fractional quantum anomalous Hall effect (FQAHE) is a fascinating emergent quantum state characterized by fractionally charged excitations in the absence of magnetic field, which could arise from the intricate interplay between electron correlation, nontrivial topology and spontaneous time-reversal symmetry breaking. 
Recently, FQAHE has been realized in aligned rhombohedral pentalayer graphene on BN superlattice (aligned R5G/BN)\cite{Julong_Nature2024}, where the topological flat band is modulated by the moir\'e potential. However, intriguingly, the FQAHE is observed only when electrons are pushed away from the moir\'e interface\cite{Julong_Nature2024,Lu_arxiv2025,Yankowitz_PRX2024}. The apparently opposite implications from these experimental observations, along with different theoretical models\cite{Jennifer_PRL2024,Senthil_PRL2024,DongJunkai_PRL2024,YahuiZhang_PRL2024,Daniel_PRB2024,Senthil_PRB2024,Devakul_PRX2024,Devakul_PRB2024,Yu_PRB2024,JianpengLiu_PRB2024,Bernevig_arXiv2023,Bernevig_arXiv2024,Zhangfan_PRB2024,Xiao_PRB2025}, have sparked intense debates regarding the role of the moir\'e potential. Unambiguous experimental observation of the topological flat band as well as moir\'e bands with energy and momentum resolved information is therefore critical to elucidate the underlying mechanism. Here by performing nanospot angle-resolved photoemission spectroscopy (NanoARPES) measurements, we directly reveal the topological flat band electronic structures of R5G, from which key hopping parameters essential for determining the fundamental electronic structure of rhombohedral graphene are extracted. 
Moreover, a comparison of electronic structures between aligned and non-aligned samples reveals that the moir\'e potential plays a pivotal role in enhancing the topological flat band in the aligned sample. Our study provides experimental guiding lines to narrow down the phase space of rhombohedral graphene, laying an important foundation for understanding exotic quantum phenomena in this emerging platform.
\end{abstract}

\newpage

\renewcommand{\thefigure}{\textbf{Fig. \arabic{figure} $\bm{|}$}}
\renewcommand{\thetable}{\textbf{Table \arabic{table} $\bm{|}$}}
\setcounter{figure}{0}

 The intricate interplay between electron correlation, band topology and spontaneous time-reversal symmetry breaking could give rise to emergent phenomena in quantum materials, e.g., fractional quantum anomalous Hall effect (FQAHE). Such FQAHE has been long sought-after\cite{Mudry_PRL2011,Wenxiaogang_PRL2011,Bernevig_PRX2011,Sheng_NC2011,Sarma_PRL2011}, not only because of the intriguing physics involving exotic fractional charge excitations in zero magnetic field\cite{MacDonald_NatRevPhys2024}, but also for potential applications in topological quantum computation\cite{DasSarma_RevModPhys2008}. Recently, FQAHE has been reported in twisted MoTe$_2$ bilayers\cite{Xu_Nature2023a,Shanjie_Nature2023,Xu_Nature2023b,Li_PRX2023} and rhombohedral pentalayer graphene aligned  on BN substrate (aligned R5G/BN)\cite{Julong_Nature2024}. 
In aligned R5G/BN, the electron correlation and nontrivial topology arise from the intrinsic topological flat band of rhombohedral graphene, whose interplay with the moir\'e potential is critical for understanding the physics of FQAHE, however, the role of the moir\'e potential has been highly debated. 
On one hand, FQAHE has been experimentally observed only in ``aligned'' rhombohedral graphene (RG) on BN substrate\cite{Julong_Nature2024,Lu_arxiv2025,Yankowitz_PRX2024} which all share a large moir\'e superlattice period $\lambda_m$ $\geq$ 10 nm, suggesting that the moir\'e potential is likely essential. On the other hand, while FQAHE\cite{Julong_Nature2024,Lu_arxiv2025,Yankowitz_PRX2024} and a rich variety of  phases, such as the extended quantum anomalous Hall effect (EQAHE)\cite{Julong_Nature2025} and superconductivity\cite{Young_Nature2025}, have been discovered in the moiré-distant regime when electrons are pushed away from the moir\'e interface, rather than in the moiré-proximal regime\cite{Ashoori_arXiv2024}, raising puzzles about the actual role of the moiré potential. Theoretical calculations adopting various interaction
schemes have provided differing views on the moir\'e potential, such as interaction-driven anomalous Hall crystal with spontaneous symmetry breaking\cite{DongJunkai_PRL2024,YahuiZhang_PRL2024,Senthil_PRL2024,Jennifer_PRL2024,Daniel_PRB2024,Senthil_PRB2024,Devakul_PRX2024}, fractional Chern insulator driven by the moir\'e potential\cite{Yu_PRB2024,JianpengLiu_PRB2024,Bernevig_arXiv2023}.  
Direct experimental observation of the topological flat band in R5G and its interplay with moir\'e bands is therefore critical for resolving puzzles regarding the moir\'e potential and for providing a more complete understanding of the fundamental physics of this FQAHE material platform.

Here by performing nanospot angle-resolved photoemission spectroscopy (NanoARPES, Fig.~1a) measurements on specially designed aligned and non-aligned samples, we directly probe the topological flat band in R5G and reveal the effect of the moir\'e potential in aligned R5G/BN sample. Thanks to the high-quality NanoARPES data, key hopping parameters are extracted by fitting the experimental data with a tight-binding model, laying a solid foundation for describing the electronic structure of RG. More importantly, we find that the aligned R5G/BN shows a stronger flat band intensity, suggesting that the topological flat band on the top surface is enhanced by the moir\'e potential on the backside of the sample. Our results highlight the pivotal role of the moir\'e potential in modifying the topological flat band in R5G/BN, which is critical for the emergent correlated electronic states.

\begin{figure*}[htbp]
	\centering
	\includegraphics[width=16.8 cm]{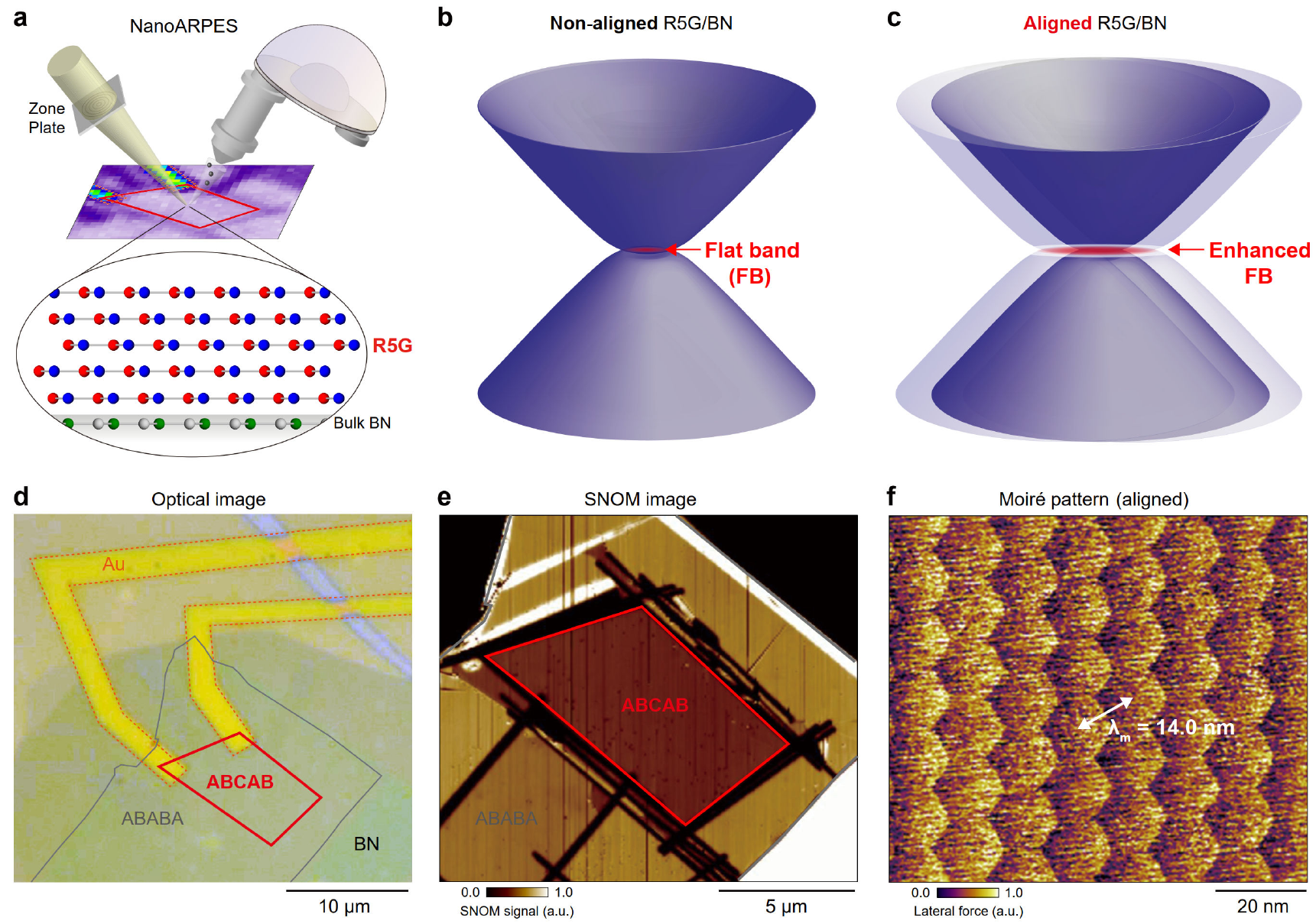}
	\caption{\textbf{Schematics and characterization of R5G/BN heterostructures (sample S1).} 
	\textbf{a}, Schematic illustration of NanoARPES experiments. \textbf{b,c}, Schematic illustration of flat bands in non-aligned (\textbf{b}) and aligned (\textbf{c}) R5G/BN. \textbf{d}, Optical image of aligned R5G/BN sample S1, where RG, Bernal stacking graphene and Au electrodes are marked by red, gray, and orange curves. \textbf{e}, SNOM image, where the dark brown region corresponds to RG. \textbf{f}, Lateral-force AFM image reveals the moir\'e period $\lambda_m$ = 14.0 $\pm$ 0.3 nm for aligned R5G/BN. }
\end{figure*}

{\bf Construction and characterization of aligned and non-aligned R5G/BN samples}

The low-energy electronic structure of few layer RG is characterized by two touching flat bands with dispersion $E\approx p^N$ where $N$ is the number of layers\cite{Koshino_PRB2009,MacDonaldPRB2010}, as schematically illustrated in Fig.~1b. In the bulk limit (infinite $N$), the flat band forms the ``drumhead'' topological surface state, which is protected by the topological properties of Dirac nodal lines of the bulk states\cite{PNAS}. 
For few layer RG, while the electronic structure of trilayer RG has been reported\cite{StarkePRB2013,BaoNL2017,CharlieNC2020}, clear experimental observation of the flat band electronic structure of R5G, especially a proper determination of its key hopping parameters, is still lacking. Moreover, revealing the effect of the moir\'e potential on the topological flat band in aligned R5G/BN is even more complicated and challenging. In principle, the introduction of a moir\'e potential with period $\lambda_m$ in aligned R5G/BN superlattice not only induces moir\'e bands which are displaced in the momentum by $k_m=2\pi/\lambda_m$, but also  might lead to modification of the flat band through the intricate interaction (Fig.~1c). 

In order to experimentally resolve the electronic structure of R5G and to evaluate the effect of moir\'e potential on the flat band electronic structure, we have constructed two types of delicately-designed samples for NanoARPES measurements --- non-aligned R5G/BN samples and aligned R5G/BN with a large moir\'e period of 14.0 nm.  
Figure~1d-e shows characterizations of the high-quality aligned R5G/BN sample S1. The R5G graphene is isolated from Bernal stacking graphene (BG) to avoid relaxation into BG during sample transfer and thermal annealing (see Extended Data Fig.~1), and connected to Au electrodes (see optical image in Fig.~1d) for NanoARPES measurements. The rhombohedral stacking is confirmed by scanning near-field optical microscopy (SNOM) measurements\cite{Keilmann_2004,Guorui_NatNano2024} shown in Fig.~1e. The good alignment between R5G and BN substrate is further supported by lateral-force atomic force microscopy (L-AFM) measurement shown in Fig.~1f (Extended Data Fig.~2 for more data), which shows a clear moir\'e superlattice with a uniform period of $\lambda_m$ = 14.0 $\pm$ 0.3 nm, comparable to that of aligned R5G/BN exhibiting FQAHE effect\cite{Julong_Nature2024,Lu_arxiv2025}.

\begin{figure*}[htbp]
	\centering
	\includegraphics[width=16.8cm]{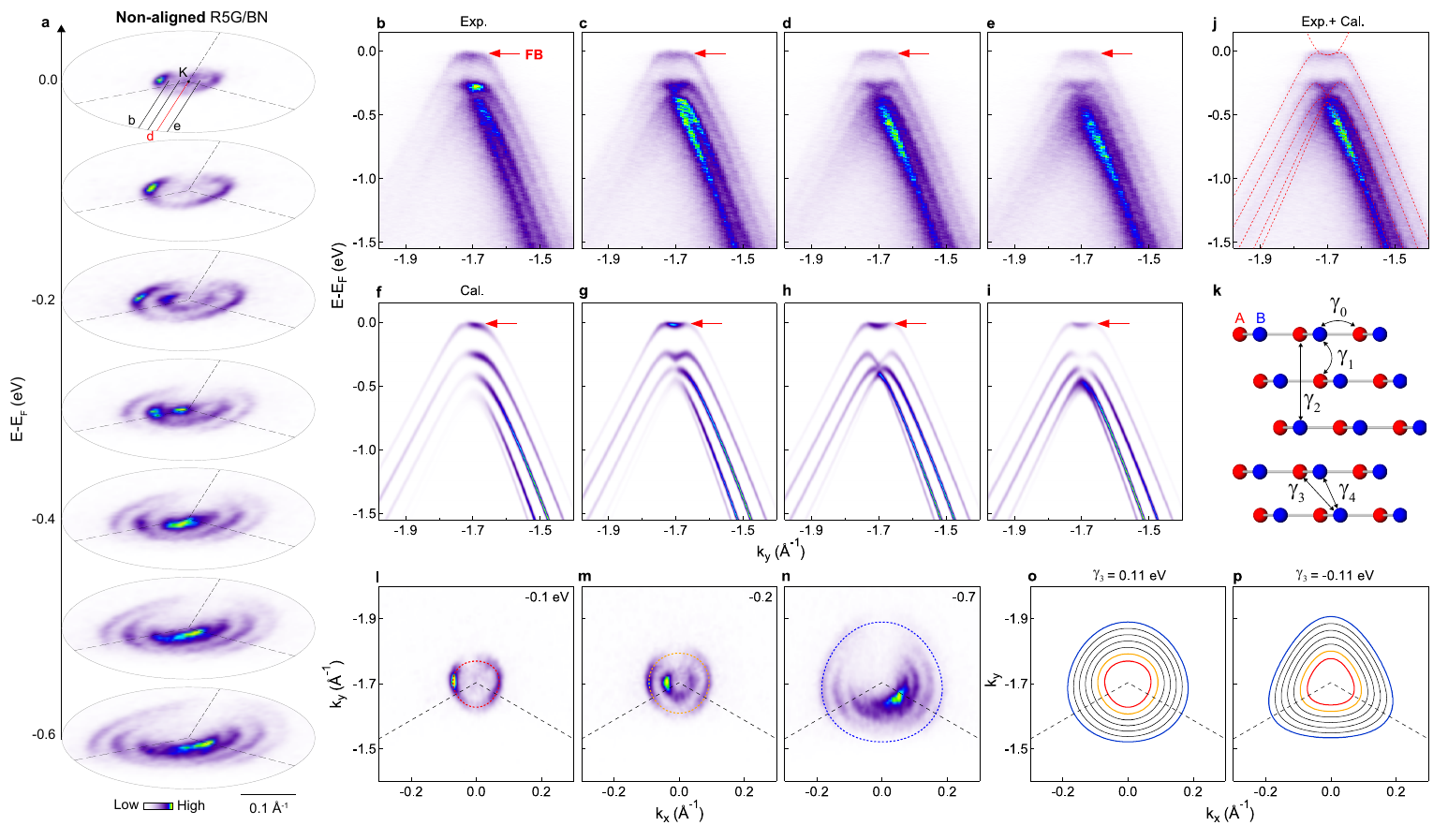}
	\caption{\textbf{Experimental electronic structure and flat band of non-aligned R5G/BN (sample S2 and S3).} 
	\textbf{a}, Intensity maps measured at energies from $E_F$ to -0.6 eV on sample S2, with photon energy of 75 eV under left circular (LC) polarization. \textbf{b-e}, Dispersion images measured on sample S3 along momentum cuts near the K point, as indicated by black and red lines in \textbf{a}. The photon energy is 95 eV under linear horizontal (LH) polarization. \textbf{f-i}, Calculated dispersion images corresponding to \textbf{b-e}.
	\textbf{j}, Comparison of dispersion image measured through the K point (\textbf{d}) and calculated dispersion (\textbf{h}, red broken curves). \textbf{k}, Atomic structure of R5G, where hopping parameters $\gamma_0$ to $\gamma_4$ are labeled. \textbf{l-n}, Intensity maps measured at energies of -0.1, -0.2 and -0.7 eV on sample S2. Red, orange and blue dotted curves are calculation results to show the reverse of warping direction. \textbf{o,p}, Calculated energy contours of the top valence band at energies from -0.1 to -0.7 eV for $\gamma_3$ = 0.11 eV and $\gamma_3$ = -0.11 eV, respectively. Red, orange and blue curves correspond to energies of \textbf{l-n}.}
\end{figure*}

{\bf Flat band electronic structure of non-aligned R5G/BN and key  hopping parameters}

We first focus on the electronic structure of non-aligned R5G/BN (sample S2 and S3 with twist angles of 36$^\circ$ and 42$^\circ$, respectively; see Extended Data Fig.~3) to reveal the intrinsic electronic structure of R5G. Figure~2a shows an overview of the electronic structure near the K point, which is distinct from the pentalayer BG (Extended Data Fig.~4). The Fermi surface map of R5G shows a small pocket centered at the K point. When moving down in energy, the pocket expands, and more pockets from other valence bands are observed. Figure~2b-e shows dispersion images measured by cutting parallel to the $\Gamma$-K direction (indicated by black and red lines in Fig.~2a), which clearly shows the flat band at the Fermi energy (indicated by red arrows). NanoARPES data measured at different sample positions show the same dispersion (Extended Data Fig.~5), again supporting that the sample is of high quality with uniform electronic structure. We note that these two non-aligned samples with different large twist angles both show the same electronic structure (see Extended Data Fig.~6 the comparison between two non-aligned samples), supporting that the experimental electronic structure is intrinsic to R5G. 

The high-quality data allow to reveal the evolution of the top valence band as well as the other four valence bands at high binding energy, which are all well captured by tight-binding calculations shown in Fig.~2f-i.  A detailed comparison between experimental data and calculated band structures (Fig.~2j) allows us to extract key hopping parameters (labeled in Fig.~2k), which are critical in determining the fundamental electronic structure of RG. In particular, NanoARPES intensity maps allow to resolve the trigonal warping in Fig.~2l-n, where the warping direction reverses from -0.1 eV (Fig.~2l) to -0.7 eV (Fig.~2n). Besides the leading hopping terms, $\gamma_0$ and $\gamma_1$, a comparison of the extensive experimental data set with theoretical calculations (Fig.~2o,p) also helps to nail down two delicate while fundamentally important inter-layer skewed hopping parameters, $\gamma_3$ and $\gamma_4$, as listed in Table 1. Note that the delicate electronic structure near the charge neutral point (CNP) is sensitively dependent on $\gamma_3$ and $\gamma_4$, which could drive Lifshitz transition\cite{Koshino_PRB2009,LauNatPhys2011} and lead to redistribution of the quantum geometry\cite{Koshino_PRB2009,Cano_PRX2025} -- one of the important ingredients for FQAHE. These extracted hopping parameters provide a reliable description of the single-particle dispersion of RG, laying an important foundation for further evaluating the effects of the moir\'e potential as well as many-body electron correlation in the correlated phenomena in aligned RG/BN.

\begin{center}
	\begin{tabular}{ c|cccc }
		\hline
		Hopping Parameters & $\gamma_0$ & $\gamma_1$ & $\gamma_3$ & $\gamma_4$\\
		\hline
		Extracted Value (eV) & $2.96$ & $0.40$ & $0.11$ & $0.14$ \\
		Uncertainty (eV) & $\pm ~0.05$ & $\pm ~0.02$ & $\pm ~0.03$ & $\pm ~0.02$\\
		\hline
	\end{tabular}
	\captionof{table}{Extracted key hopping parameters from R5G. Here we extract only the in-plane and out-of-plane nearest-neighbor hopping parameters. Note that the out-of-plane second-nearest-neighbor hopping $\gamma_2$ is on the order of 10 meV\cite{Koshino_PRB2009}, which is too small to be extracted from our data.}
\end{center}

\begin{figure*}[htbp]
	\centering
	\includegraphics[width=16.8 cm]{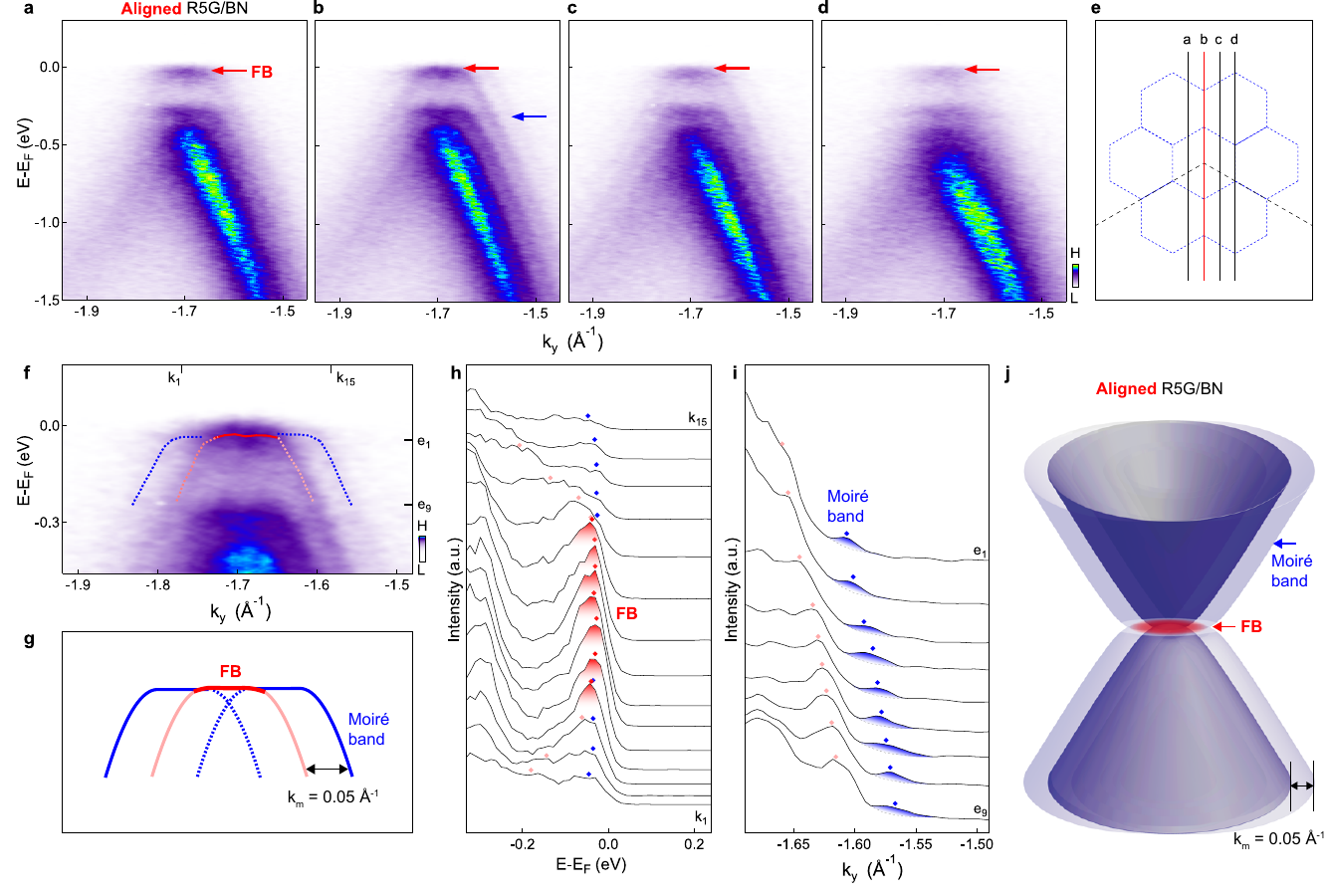}
	\caption{\textbf{Observation of strong flat band and moir\'e replica bands in aligned R5G/BN (sample S1).} \textbf{a-d}, Dispersion images measured along momentum lines parallel to $\Gamma$-K direction, as indicated by black and red lines in (\textbf{e}). Red and blue arrows indicate the flat band and weak moir\'e band. \textbf{f}, Zoom-in dispersion image through the K point (data shown in \textbf{b}) using a log color scale. \textbf{g}, Schematic dispersion of flat band (red curve) and moir\'e bands (blue curves). \textbf{h}, EDCs extracted at momenta from $k_1$ to $k_{15}$ as indicated in (\textbf{f}). Red and blue dots indicate peaks from the flat band and moir\'e replica bands. \textbf{i}, MDCs extracted at energy from $e_1$ to $e_9$ as indicated in (\textbf{f}). \textbf{g}, Schematic summary of the experimental electronic structure of aligned R5G/BN.}
	\label{Fig3}
\end{figure*}

{\bf Observation of enhanced flat band in aligned R5G/BN}

Figure~3 shows the electronic structure of aligned R5G/BN sample S1. A strong flat band is clearly resolved in the dispersion images measured parallel to the $\Gamma$-K direction (Fig.~3a-d, pointed by red arrows), which is clearly different from data presented above on non-aligned R5G/BN. Moreover, clear moir\'e bands are captured in Fig.~3b (indicated by blue arrow), which are better resolved in the zoom-in image in Fig.~3f. We would like to point out that ARPES is an extremely surface-sensitive probe\cite{ShenRMP2003,ZhangHYPrimers} with the strongest contribution from the top layer, and therefore the clear observation of moir\'e bands in the NanoARPES data here suggests that the top R5G surface, which is far away from the moir\'e interface on the backside of R5G, is also modulated by the moir\'e potential.

Figure 3g-i further shows a quantitative analysis of the flat band and moir\'e bands. The flat band is clearly identified in the energy distribution curves (EDCs) shown in Fig.~3h (red shaded peaks), while the moir\'e bands are better resolved in the momentum distribution curves (MDCs) shown in Fig.~3i (blue shaded peaks). Combining EDC and MDC analysis, the extracted dispersions for the flat band and moir\'e bands are overplotted in Fig.~3f, where the momentum displacement of the moir\'e bands by $k_m$ = 0.05 $\pm$ 0.01 $\text{\AA}^{-1}$ (Fig.~3g) is in good agreement with the moir\'e period $\lambda_m$ = 14.0 nm (Fig.~1f). We note that the momentum range of the flat band ($2p_0$) in bulk RG\cite{PNAS} is determined by the ratio between the out-of-plane ($\gamma_1$) and in-plane ($\gamma_0$) nearest hopping parameters by $2p_0 = 4 \gamma_1/(\sqrt{3}a\gamma_0)$, where $a$ is the lattice constant. Here the experimentally extracted  momentum range of the flat band $\Delta k$ = 0.10 $\pm$ 0.02 $\text{\AA}^{-1}$ in R5G (see Extended Data Fig.~7 for more details) approaches that in the bulk rhombohedral graphite\cite{PNAS}. The momentum range in which the flat band spans is larger than the moir\'e superlattice vector $k_m$ = 0.05 $\text{\AA}^{-1}$, ensuring that the moir\'e bands overlap with the flat band of R5G. Such overlapping is critical for obtaining an isolated flat band and enhancing the flat band, as schematically illustrated in Fig.~3j.

\begin{figure*}[htbp]
	\centering
	\includegraphics[width=16.8cm]{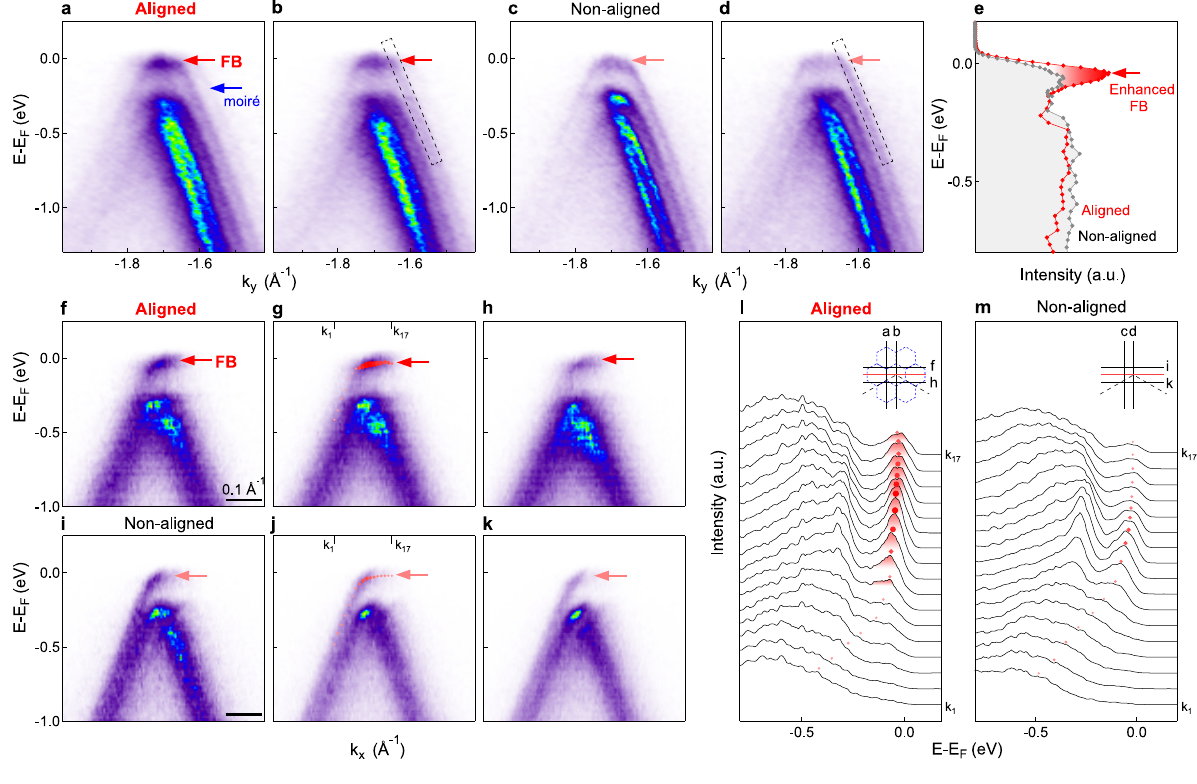}
	\caption{\textbf{Stronger flat band in the aligned  R5G/BN sample as compared to the non-aligned sample.} \textbf{a,b}, Dispersion images of aligned R5G/BN sample (S1) measured parallel to $\Gamma$-K direction (indicated by vertical lines in \textbf{l}). \textbf{c,d}, Similar data to \textbf{a,b} measured on non-aligned R5G/BN sample (S3). \textbf{e}, Comparison of EDCs on aligned and non-aligned samples, obtained by integrating over boxes marked in \textbf{b} and \textbf{d}. \textbf{f-h}, Dispersion images measured along momentum cuts indicated by horizontal lines in \textbf{l} on aligned sample S1. \textbf{i-k}, Dispersion images measured along momentum cuts in \textbf{m} on non-aligned sample S3. \textbf{l,m}, EDC analysis for the aligned and non-aligned samples, respectively. The EDCs are extracted from momentum range of $k_1$ to $k_{17}$ as indicated by black ticks in (\textbf{g,j}). Red dots in (\textbf{g,j,l,m}) indicate the extracted dispersion with the color and size representing the peak intensity.}
	\label{Fig4}
\end{figure*}

{\bf Comparison between aligned and non-aligned R5G/BN samples}

To further reveal the effect of the moir\'e potential, we show in Fig.~4 a side-by-side comparison of experimental electronic structures between aligned and non-aligned R5G/BN samples. Important to note that in order to have a fair comparison, here the measurements were performed at the same endstation under the same experimental conditions. The moir\'e bands are observed in the aligned sample in Fig.~4a,b (pointed by blue arrow) when cutting parallel to the $\Gamma$-K direction, while absent in the non-aligned sample (Fig.~4c,d). More importantly, the flat band intensity is strongly enhanced in the aligned sample (indicated by red arrow in Fig.~4a,b) as compared to the non-aligned sample (Fig.~4c,d), which is supported by the intensity curve shown in Fig.~4e (see more comparison in Extended Data Fig.~8). The stronger flat band is also clearly revealed when comparing dispersion images measured by cutting perpendicular to the $\Gamma$-K direction (pointed by red arrows in Fig.~4f-h and Fig.~4i-k). In particular, in the EDC stacks in Fig.~4l, the flat band in the aligned sample  shows a much stronger intensity near $E_F$ (highlighted by red dots and shadings), while in the non-aligned sample, weaker peaks are only barely detectable near $E_F$  (Fig.~4m). The enhancement of the flat band intensity clearly demonstrates that the moir\'e potential does play an important role in modifying the flat band electronic structure of aligned R5G/BN, even though the moir\'e interface on the backside is far away from the top surface exposed in NanoARPES measurements.

The observation of clear moir\'e bands and stronger flat band in aligned R5G/BN suggests that electrons located on the top surface (moir\'e-distant surface) experience a spatially modulated potential, whose periodicity follows that of the moir\'e superlattice. We note that in various theoretical calculations\cite{Yu_PRB2024,YahuiZhang_PRL2024,DongJunkai_PRL2024,Zhangfan_PRB2024,JianpengLiu_PRB2024,Xiao_PRB2025}, a moir\'e potential of no more than 30 meV is often applied only to the moir\'e-proximal surface, and its main effect is to gap or isolate the topological flat band from other bands (Extended Data Fig.~9).  Although such gap opening is beyond the state-of-art NanoARPES experimental resolution, the mini-gap opening could be part of the reason for the enhancement of the flat band intensity in our experimental results. However, other than this, the effect of moir\'e potential on the electronic structure is smaller than our experimental results. For example, in the calculated electronic structure (Extended Data Fig.~9), moir\'e bands outside R5G bands are barely discernible even when using a moir\'e potential energy of 30 meV, again suggesting that the moir\'e potential in the calculations is underestimated.  How to properly incorporate the synergistic effect of moir\'e potential into the physics of aligned R5G/BN would require extensive theoretical investigations, and here our experimental results provide useful benchmarks to testify different theoretical models.

{\bf Conclusion and outlook}

To summarize, through high-resolution NanoARPES measurements and  careful comparison of experimental results with theoretical calculations, we extract important hopping parameters which are essential for determining the fundamental electronic structure of R5G, including the higher order hopping parameters whose values sensitively determine the trigonal warping and Fermi surface topology etc.  Moreover, the observation of moir\'e bands and enhanced flat band intensity highlights the important role of moir\'e potential in the aligned R5G/BN. 
Considering that a variety of novel quantum phases including FQAHE have been reported only when RG is subjected to a moir\'e superlattice\cite{WangFTrilayerNat19FM,Julong_NatNano2024,Guorui_Science2024,Lu_arxiv2025,Yankowitz_PRX2024,Ashoori_arXiv2024,Guorui_arXiv2024}, how the moir\'e  potential affects the intricate topology property of the flat band, the charge modulation, and the gap in aligned R5G/BN is critical for understanding the underlying physics.  Our work calls attention to the strong effect of moir\'e potential on the electronic structure of aligned R5G/BN even in the moir\'e-distant regime, which could help to reconcile different  experimental observations and theoretical models.

\newpage

\begin{methods}

\renewcommand{\thefigure}{\textbf{Extended Data Fig.~\arabic{figure} $\bm{|}$}}
\setcounter{figure}{0}

\subsection{Sample preparation and characterization.}

Few-layer RG flakes were mechanically exfoliated from natural graphite onto SiO$_2$/Si substrates, and their thickness was determined by optical reflectance contrast and AFM measurement. The RG domains were identified by using SNOM or infrared imaging, and subsequently isolated from the surrounding BG by AFM tip\cite{Shi_NanoLett2018}. We sequentially picked up the BN flake and R5G using polypropylene carbonate (PPC) film. During this process, the straight edges of the R5G flakes were intentionally aligned at specific angles with respect to the bottom BN to form a moir\'e superlattice or non-aligned sample.  The R5G/BN was then flipped and transferred onto a clean SiO$_2$/Si substrate, which was then annealed at 350$~^\circ$C in vacuum to remove the PPC film underneath the heterostructure. Metal electrodes were deposited following standard e-beam lithography.  During the entire sample preparation process, the rhombohedral stacking was frequently checked and verified by SNOM measurements to ensure that the stacking order remains the same.

\subsection{ARPES measurements.}

NanoARPES measurements were performed at beamline ANTARES of the Synchrotron SOLEIL in France and beamline I05 of the Diamond Light Source in the UK with beam sizes smaller than $\SI{1}{\micro\meter}$. Before NanoARPES measurements, the few-layer RG samples were annealed at 180$~^\circ \text{C}$ for several hours in ultrahigh vacuum (UHV) until sharp dispersions were observed.  Sample regions with different stacking orders were directly distinguished from the experimental spectra (Extended Data Fig.~4), and the homogeneity across the sample was checked (Extended Data Fig.~5). 

Data shown in Figure~2a,l,m,n were measured at Diamond Light Source at a temperature of 25 K using a photon energy of 75 eV with LC polarization, with total energy resolution better than 33 meV. Other data were measured at SOLEIL at temperature of 78 K using a photon energy of 95 eV with LH polarization, with energy resolution of 40 meV.

\subsection{Twist angle and electronic structure of two non-aligned R5G/BN samples.}

The twist angle of non-aligned R5G/BN samples are determined both from the real-space armchair directions and momentum-space K points of R5G and BN as shown in Extended Data Fig.~3. Here the twist angle of non-aligned sample S2 and sample S3 are determined to be 36$^\circ$ $\pm$ 1$^\circ$ and 42$^\circ$ $\pm$ 2$^\circ$. Extended Data Fig.~6 shows the comparison of dispersion images measured on sample S2 and sample S3, showing the same electronic structure intrinsic to R5G despite the different twist angles.

\subsection{Momentum range of the flat band in non-aligned R5G/BN.}

Extended Data Fig.~7 shows the Fermi surface and dispersion images cutting along the $\Gamma$-K directions, with a flat band being clearly observed spanning a momentum range of 0.10 $\pm$ 0.02 $\text{\AA}^{-1}$ as labeled by black arrows in Extended Data Fig.~7a,c.

\subsection{Enhancement of the flat band from EDC analysis along $\Gamma$-K direction.}

Extended Data Fig.~8 shows the dispersion images cutting along the $\Gamma$-K direction and the corresponding EDC analysis. The flat band with a stronger intensity is not only directly observed from the dispersion images (Extended Data Fig.~8a,b), but also from the extracted spectral weight of flat band (Extended Data Fig.~8c,f). Here we use the color and size of red dots to represent the extracted spectral weight. The extracted spectra weight shows that the flat band has a stronger intensity in the aligned sample.

\subsection{Theoretical calculations.}

For pristine R5G, we adopt the tight-binding model\cite{Yu_PRB2024} to capture its electronic structure at one valley, 
\begin{align} 
	H_{0} &= \sum_{\mathbf{k}} \sum_{l=0}^{4} \sum_{\alpha,\alpha'} c^\dagger_{\mathbf{k},l,\alpha} \begin{pmatrix}
		V_l & \gamma_0 f(\mathbf{k}) \\
		\gamma_0f^*(\mathbf{k}) & V_l \\
	\end{pmatrix}_{\alpha,\alpha'}  c_{\mathbf{k},l,\alpha'} \\\nonumber
	&+ \sum_{\mathbf{k}} \sum_{l=0}^{3} \sum_{\alpha, \alpha'} c^\dagger_{\mathbf{k},l,\alpha} \begin{pmatrix}
		- \gamma_4 f(\mathbf{k})& - \gamma_3 f^*(\mathbf{k}) \\
		\gamma_1 & - \gamma_4 f(\mathbf{k}) \\
	\end{pmatrix}_{\alpha,\alpha'} c_{\mathbf{k},l+1,\alpha'} + \mathrm{H.c.} \\\nonumber
	&+ \sum_{\mathbf{k}} \sum_{l=0}^{2} \sum_{\alpha,\alpha'} c^\dagger_{\mathbf{k},l,\alpha} \begin{pmatrix}
		0 & \gamma_2 \\
		0 & 0 \\
	\end{pmatrix}_{\alpha,\alpha'} c_{\mathbf{k},l+2,\alpha'} + \mathrm{H.c.}
\end{align}
The index $\alpha=A,B$ denotes graphene sublattices, and $l =0, 1, \dots, 4$ denote layers from bottom to top, and the momentum $\mathbf{k}$ is measured from the $\mathbf{K}$ point of R5G Brillouin zone.
The $c^\dagger_{\mathbf{k},l,\alpha}$ ($c_{\mathbf{k},l,\alpha}$) are electron creation (annihilation) operators. The hopping parameters $\gamma_0, \gamma_1, \dots, \gamma_4$ are schematically shown in Fig.~2k. The factor $f(\mathbf{k}) \approx -\frac{\sqrt{3}}{2} a_0 k_- + \frac{1}{8} (a_0 k_+)^2$ expands to the second order of $k$, where $k_\pm=k_x\pm ik_y$. The potential on the $l$ layer, $V_l = V_{\rm ISP} |2-l|$, where $V_{\rm ISP}$ = -49 meV describes the intrinsic inversion-symmetric potential.

For aligned R5G/BN, the BN substrate induces a moir\'e potential on the bottom layer\cite{Yu_PRB2024}, 
\begin{align}
	H_1 &= \sum_{\mathbf{k}} \sum_{\alpha}  c^\dagger_{\mathbf{k},0,\alpha} V_{0;\alpha} c_{\mathbf{k},0,\alpha'} 
	+  \sum_{\mathbf{k}} \sum_{\alpha,\alpha'} \sum_{j=1}^{3} c^\dagger_{\mathbf{k}+\mathbf{g}_j,0,\alpha} V_{1;\alpha,\alpha'} \begin{pmatrix}
		1 & \omega^{-j} \\
		\omega^{j+1} & \omega \\
	\end{pmatrix}_{\alpha,\alpha'} c_{\mathbf{k},0,\alpha'} + \mathrm{H.c.}
\end{align}
where $\omega=e^{i2\pi/3}$. $\mathbf{g}_1$,$\mathbf{g}_2$ and $\mathbf{g}_3$ denote the three moir\'e reciprocal lattice vectors. We take the zeroth and first-harmonics of the moir\'e potential $V_0$ = 20 meV and $V_1$ = 30 meV in the calculation of Extended Data Fig.~9b,d. However, neither the moir\'e bands nor the enhancement of the flat band is captured in the calculations, indicating that the moir\'e potential is still underestimated.

The NanoARPES intensity is simulated based on the Fermi golden rule, taking into account the quantum coherent interference\cite{Zhu_2021_theory,Zhou_NM2024}
\begin{align}
	I(\mathbf{p}, E)  &\propto \sum_{n} \sum_{\mathbf{k}\in\mathrm{mBZ}} \sum_{\mathbf{G}} \delta_{\mathbf{p}_\parallel , \mathbf{k}+\mathbf{G}+\mathbf{K}} \left|  \sum_{l',\alpha'} e^{-i p_z c_0 l' } \left( \mathbf{A} \cdot \mathbf{v}_{l'\alpha',l\alpha} \right)  U_{\mathbf{G},l,\alpha;n}(\mathbf{k})  \right|^2 \delta(E - E_{\mathbf{k},n}) 
\end{align}
where $U_{\mathbf{G},l,\alpha;n}(\mathbf{k})$ and $E_{\mathbf{k},n}$ are the wavefunction and energy of the initial Bloch state, $p_{||}$ and $p_z$ denote the in-plane and out-of-plane photoelectron momentum, and $c_0$ denotes the inter-layer spacing of RG which is typically 0.33 nm. An imaginary part of $p_z$ is also introduced to model the signal decay beneath the surface. $\mathbf{A}$ is the polarization vector of the incident light, and $\mathbf{v} = \nabla_{\mathrm{k}}\mathnormal{H}(\mathrm{k})$ denotes the velocity operator.

\newpage
	
\renewcommand{\thefigure}{\textbf{Extended Data Fig.~\arabic{figure} $\bm{|}$}}
\setcounter{figure}{0}

\begin{figure*}[htbp]
	\centering
	\includegraphics[width=11cm]{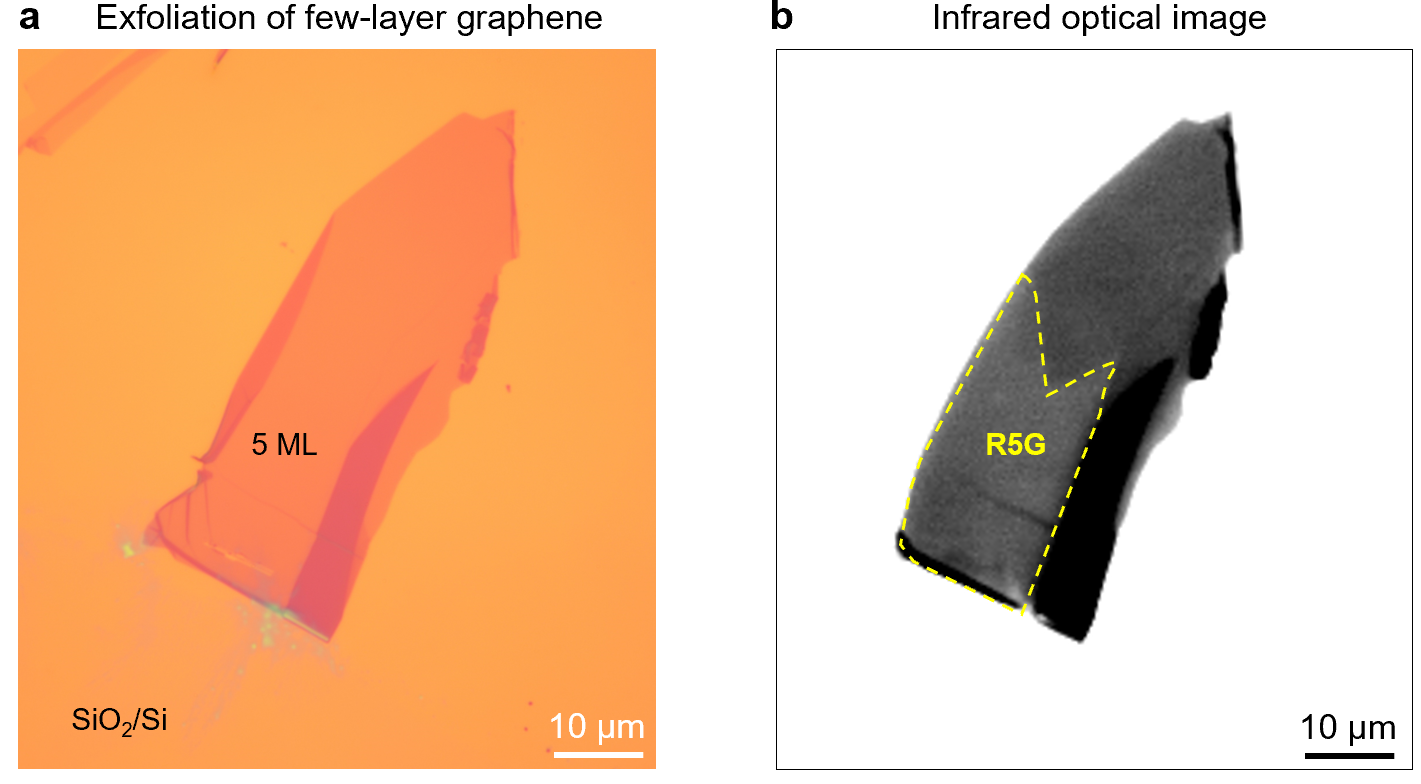}
	\caption{\textbf{Identification and isolation of R5G.} \textbf{a}, Optical image of R5G exfoliated on SiO$_2$/Si. \textbf{b}, Infrared optical image to reveal the stacking of RG (indicated by yellow curve).}
	\label{ExFig1}
\end{figure*}

\begin{figure*}[htbp]
	\centering
	\includegraphics[width=17cm]{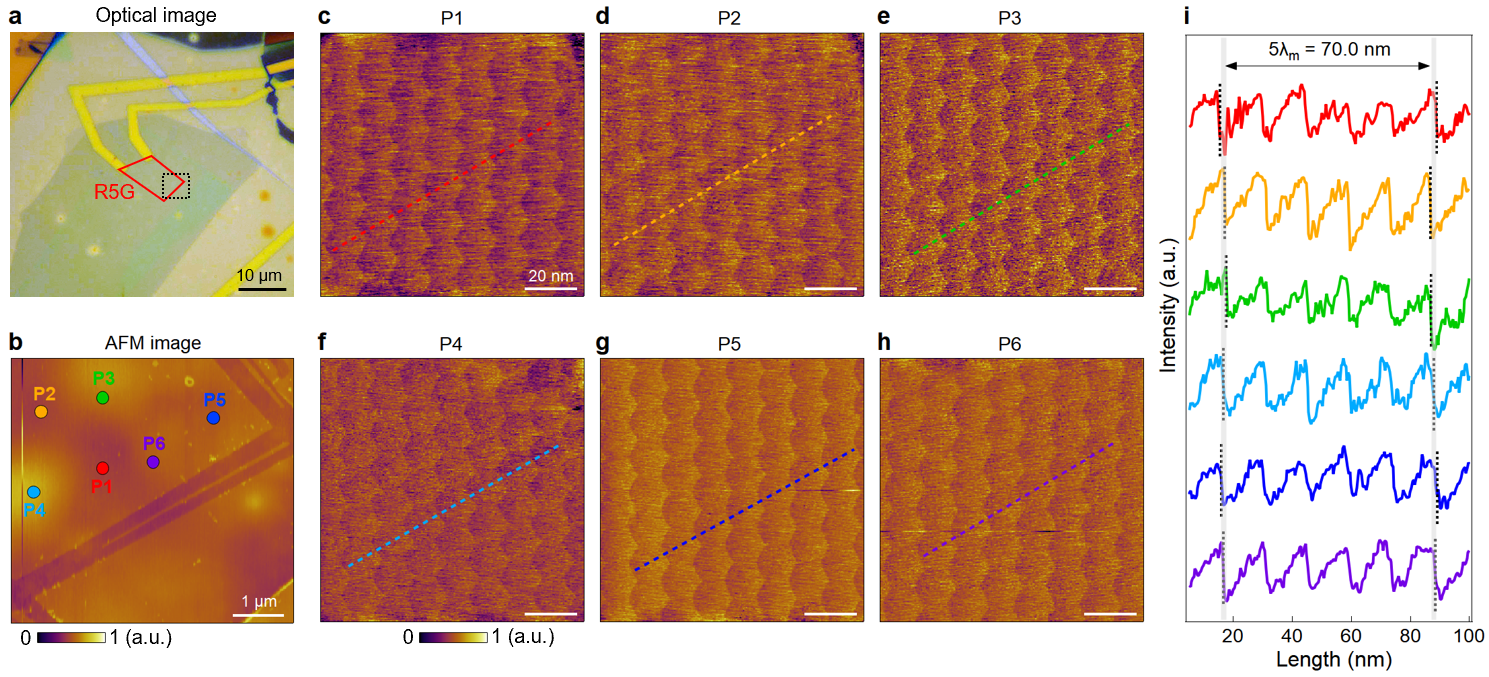}
	\caption{\textbf{Homogeneity of the moir\'e period.} \textbf{a}, Optical image of aligned R5G/BN (sample S1). The red curve indicates R5G region. \textbf{b}, AFM image by zooming in a region indicated by black dotted box in \textbf{a}. \textbf{c-h}, Lateral-force AFM images to show the moir\'e pattern measured at different positions from P1 to P6 as labeled in \textbf{b}. \textbf{i}, AFM line profiles measured along colored lines in \textbf{c-h}, with all positions share similar moir\'e period of $\lambda_m$ = 14.0 $\pm$ 0.3 nm.}
	\label{ExFig2}
\end{figure*}

\begin{figure*}[htbp]
	\centering
	\includegraphics[width=17cm]{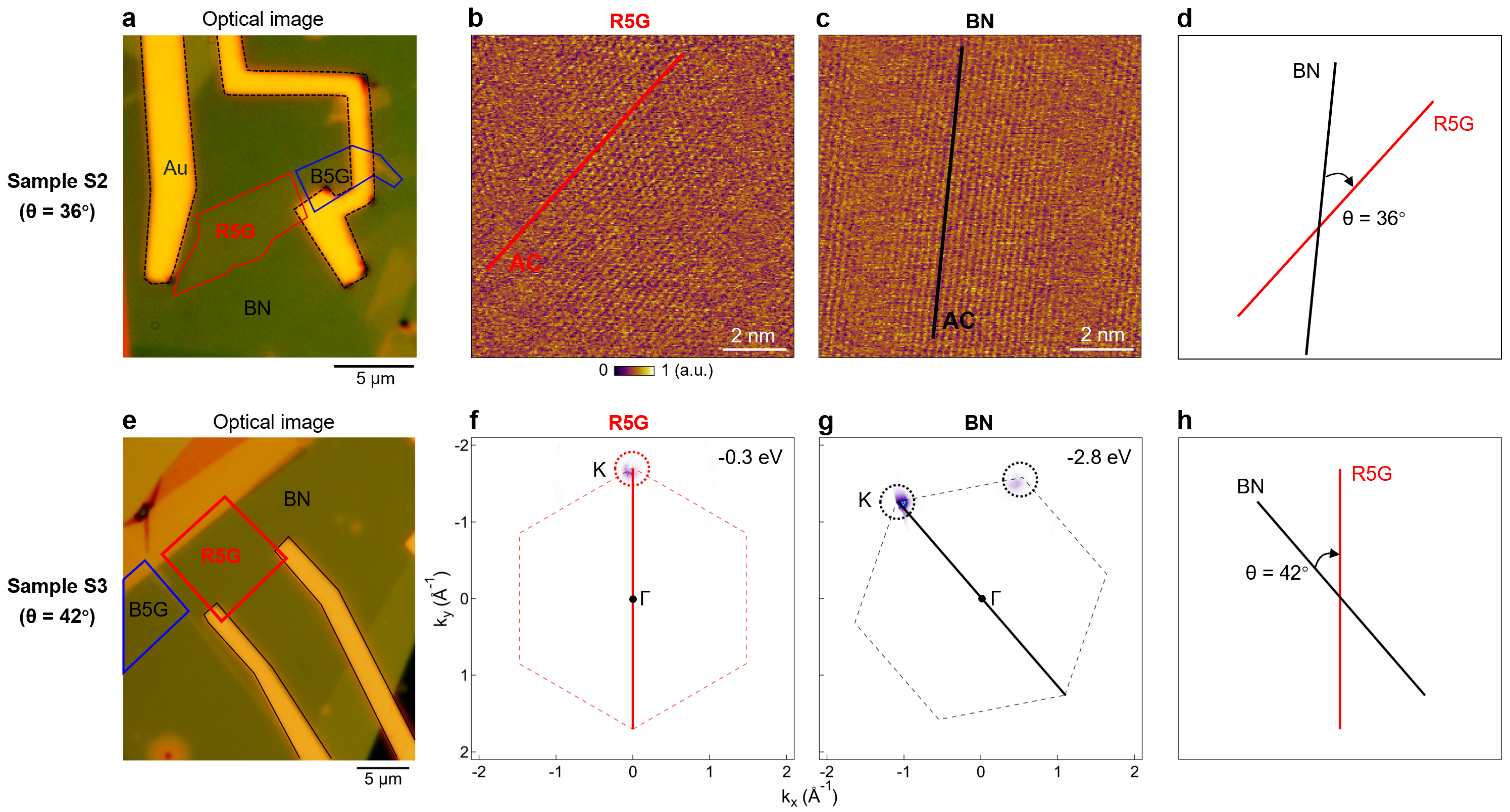}
	\caption{\textbf{Determination of twist angle for non-aligned R5G/BN samples.} \textbf{a}, Optical image of sample S2. \textbf{b,c}, AFM images measured on R5G and BN regions on sample S2, with red and black lines indicating the armchair (AC) direction. The twist angle is determined to be 36$^\circ$ $\pm$ 1$^\circ$ from BN to R5G (\textbf{d}). \textbf{e}, Optical image of sample S3. \textbf{f,g}, Intensity maps measured at energies of -0.3 eV and -2.8 eV on sample S3, with K points of R5G and BN indicated by red and black dotted circles. The twist angle is determined to be 42$^\circ$ $\pm$ 2$^\circ$ from BN to R5G (\textbf{h}).}
	\label{ExFig5}
\end{figure*}

\begin{figure*}[htbp]
	\centering
	\includegraphics[width=17cm]{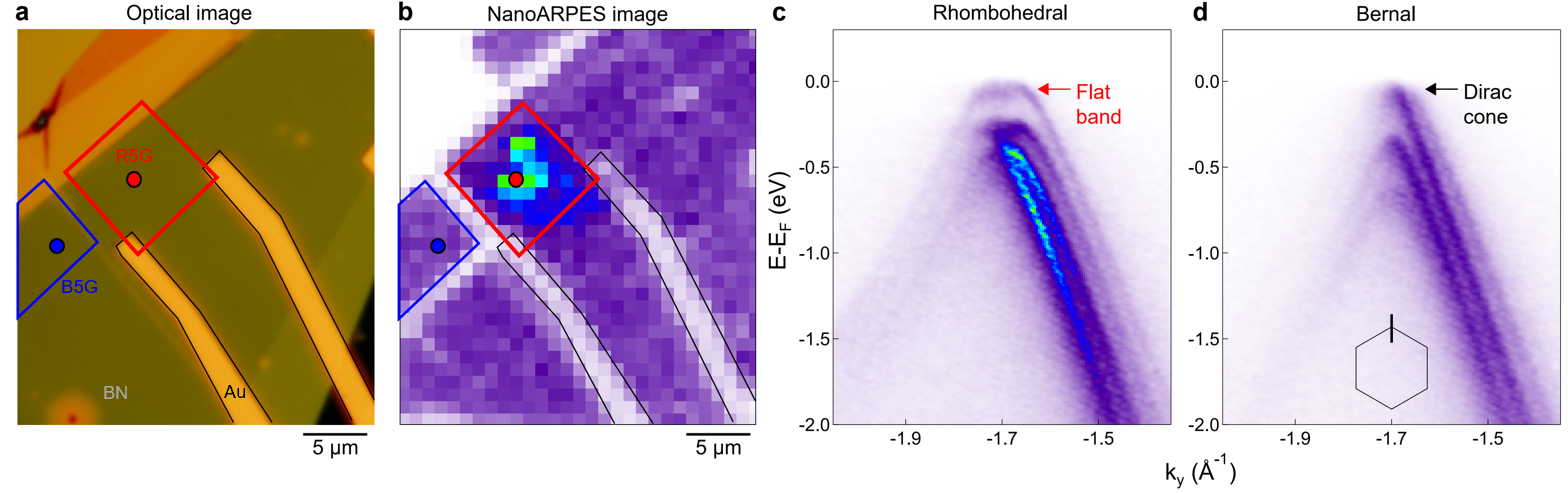}
	\caption{\textbf{Distinct electronic structure between pentalayer rhombohedral and Bernal stacking graphene.} \textbf{a}, Optical image of non-aligned R5G/BN (sample S3), with red, blue and black curves indicating R5G, pentalayer Bernal graphene (B5G) and electrodes. \textbf{b}, Spatially-resolved NanoARPES intensity map with a similar range to the optical image.  The NanoARPES intensity is obtained by integrating intensity from $E_F$ to -1.0 eV. \textbf{c,d}, Experimental dispersion images measured on R5G and B5G along $\Gamma$-K direction (vertical line in the inset).}
	\label{ExFig4}
\end{figure*}

\begin{figure*}[htbp]
	\centering
	\includegraphics[width=13cm]{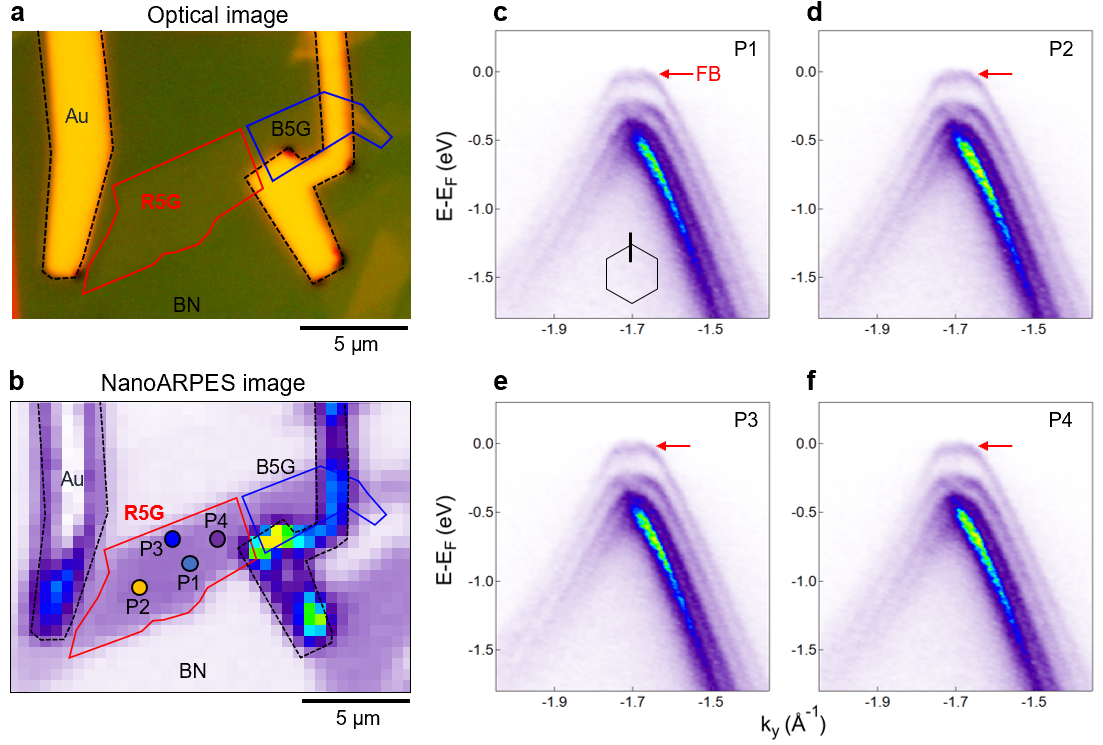}
	\caption{\textbf{Uniform electronic structure at different sample positions.} \textbf{a}, Optical image of non-aligned R5G/BN (sample S2), with red and blue curves indicating R5G and B5G regions. \textbf{b}, NanoARPES spatial image with a similar range of the optical image. \textbf{c-f}, Dispersion images measured at positions from P1 to P4 as indicated in \textbf{b}, with the cutting direction along $\Gamma$-K as indicated by vertical line in the inset.}
	\label{ExFig3}
\end{figure*}

\begin{figure*}[htbp]
	\centering
	\includegraphics[width=17cm]{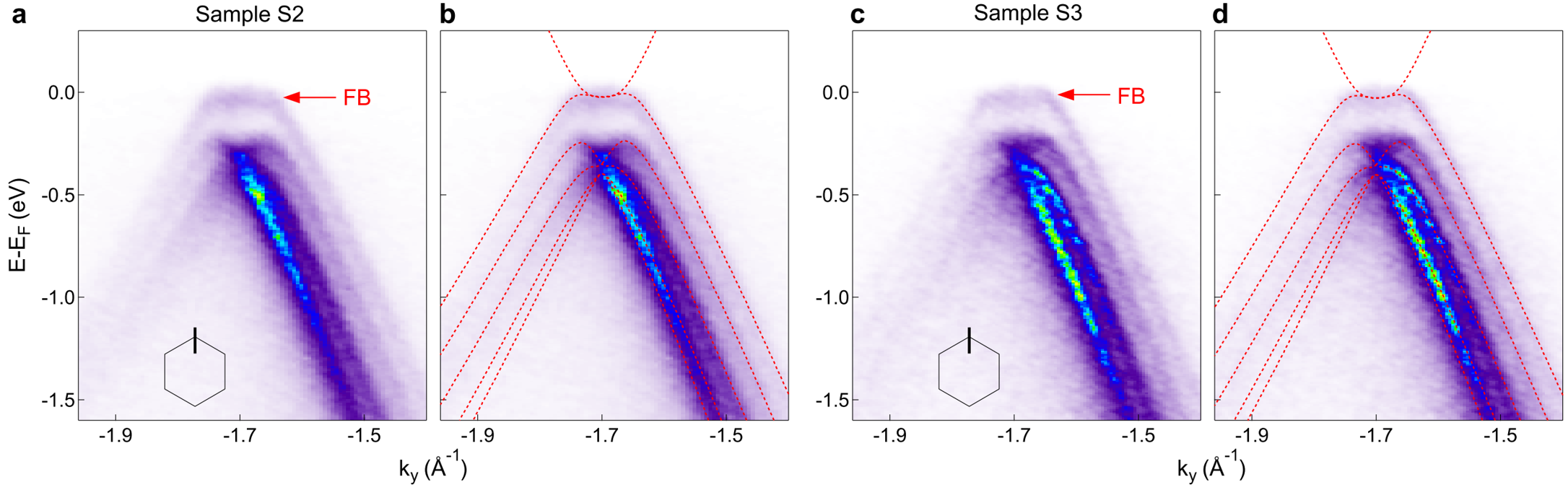}
	\caption{\textbf{Comparison of electronic structure between non-aligned R5G/BN sample S2 and S3.} \textbf{a}, Dispersion image measured along $\Gamma$-K direction on sample S2. \textbf{b}, Same image as \textbf{a} with calculated results appended as red dotted curves. \textbf{c}, Dispersion image measured along $\Gamma$-K direction on sample S3. \textbf{d}, Same image as \textbf{c} with calculated results appended as red dotted curves.}
	\label{ExFig7}
\end{figure*}

\begin{figure*}[htbp]
	\centering
	\includegraphics[width=17cm]{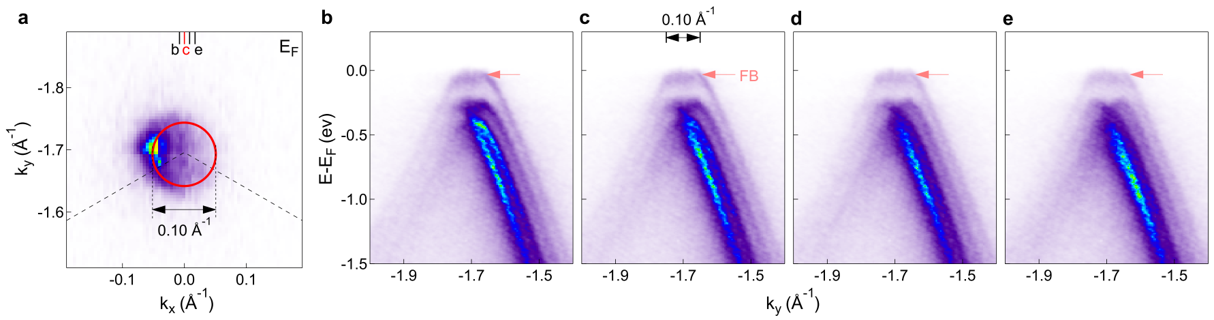}
	\caption{\textbf{Momentum range of the flat band in non-aligned R5G/BN.} \textbf{a}, Experimental Fermi surface map on the non-aligned R5G/BN (sample S3). The red circle indicates the flat band with a diameter of 0.10 $\pm$ 0.02 $\text{\AA}^{-1}$. \textbf{b-e}, Experimental dispersion images measured along directions indicated by short lines in \textbf{a}.}
	\label{ExFig9}
\end{figure*}

\begin{figure*}[htbp]
	\centering
	\includegraphics[width=17cm]{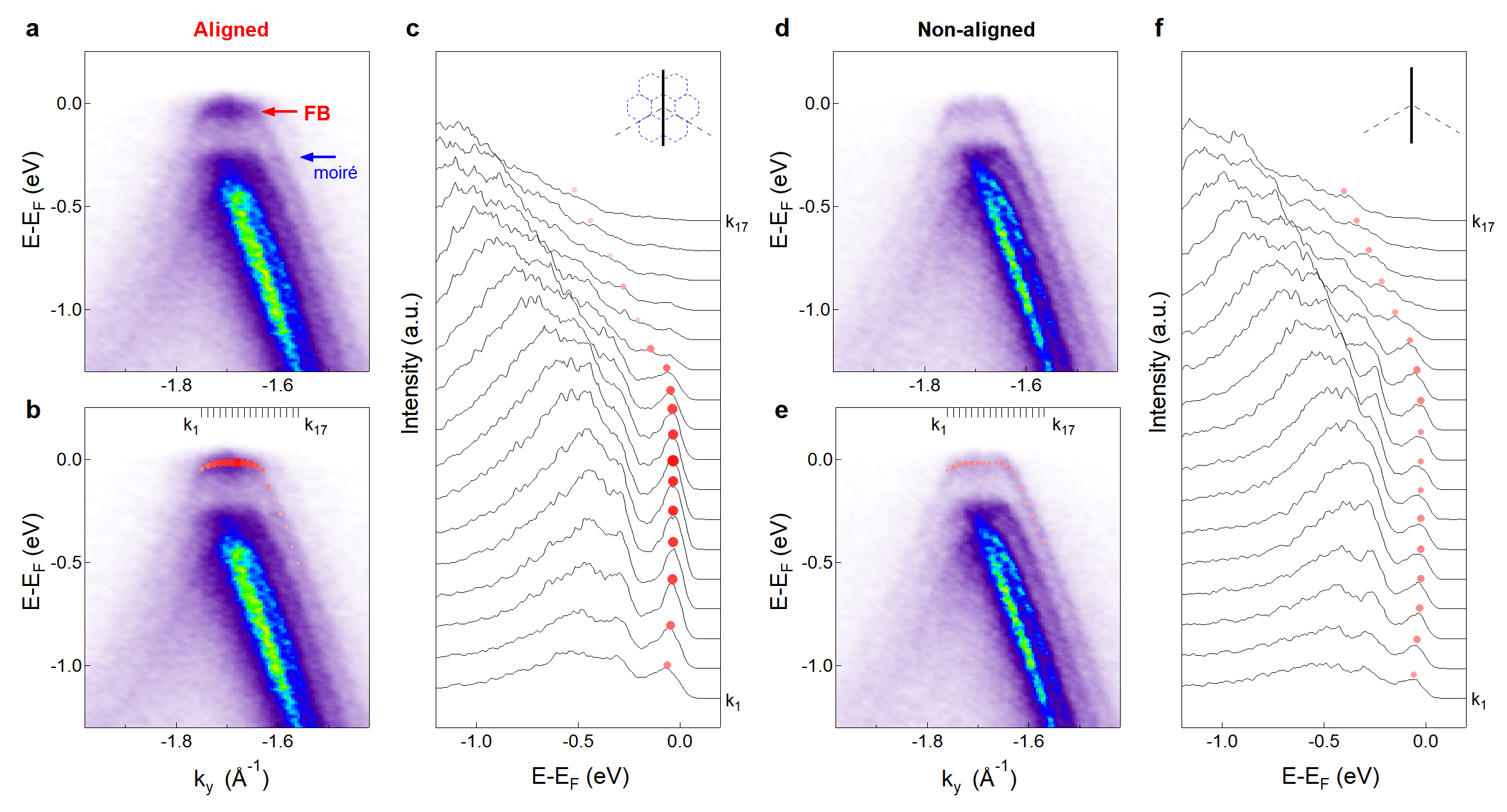}
	\caption{\textbf{Enhanced flat band in aligned R5G/BN along $\Gamma$-K direction.} \textbf{a}, Experimental dispersion image measured along $\Gamma$-K direction in aligned sample S1. \textbf{b}, Same as \textbf{a} with extracted flat band dispersion and spectral weight overplotted as red dots. \textbf{c}. Extracted EDCs from momentum as indicated by short ticks in \textbf{a}.  \textbf{d}, Experimental dispersion image measured along $\Gamma$-K direction in non-aligned sample S3. \textbf{e}, Same as \textbf{d} with extracted flat band dispersion and spectral weight overplotted as red dots. \textbf{f}, Extracted EDCs from momentum as indicated by short ticks in \textbf{e}. The color and size of red dots in all panels correspond to the spectral weight of the band.}
	\label{ExFig10}
\end{figure*}

\begin{figure*}[htbp]
	\centering
	\includegraphics[width=16cm]{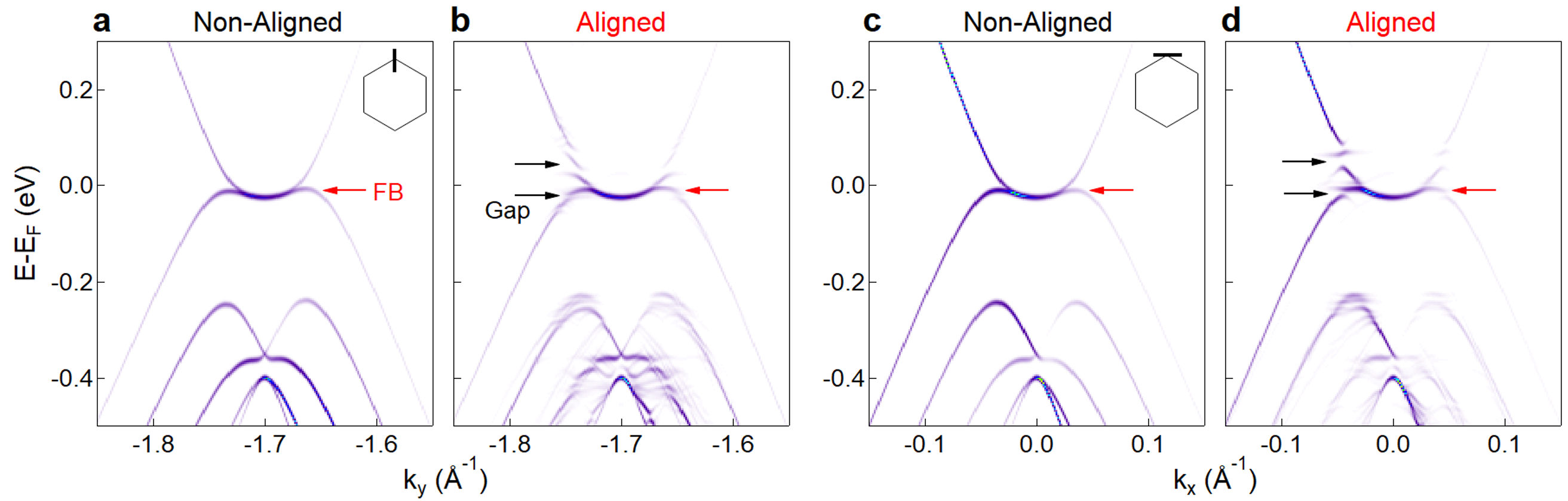}
	\caption{\textbf{Comparison of calculated electronic spectra in non-aligned and aligned samples.} \textbf{a,b}, Calculated spectrum by cutting along $\Gamma$-K direction for non-aligned and aligned R5G/BN, respectively. \textbf{c,d}, Calculated spectrum by cutting perpendicular to $\Gamma$-K direction for non-aligned and aligned R5G/BN, respectively.}
	\label{ExFig10}
\end{figure*}

\end{methods}

\newpage
\begin{addendum}
	
	\item[Data availability] The data that support the plots within this paper and other findings of this study are available from the corresponding author upon reasonable request.
	
	\item[Acknowledgement]
	This work is supported by the National Key R $\&$ D Program of China (Grant No.~2021YFA1400100), the National Natural Science Foundation of China (Grant No.~12234011, 12421004, 52388201, 92250305, 52025024, 12327805, 11725418, 12350005, 12174248), and the New Cornerstone Science Foundation through the XPLORER PRIZE. G.C. acknowledges Shanghai Science and Technology Innovation Action Plan (Grant No.~24LZ1401100) and Yangyang Development Fund. H.Z. acknowledges support from the Shuimu Tsinghua Scholar program, the National Natural Science Foundation of China (Grant No.~12304226), and the Project funded by China Postdoctoral Science Foundation (Grant No.~2022M721887). T.S. acknowledges support from JST-CREST (Grant No. JPMJCR18T1). K.L. acknowledges support from the National Natural Science Foundation of China (Grant No.~124B2071). K.W. and T.T. acknowledge support from the JSPS KAKENHI (Grant Numbers 21H05233 and 23H02052) , the CREST (JPMJCR24A5), JST and World Premier International Research Center Initiative (WPI), MEXT, Japan. 	
	We acknowledge SOLEIL for the provision of synchrotron radiation facilities of beamline ANTARES, and the Diamond Light Source for the provision of synchrotron radiation facilities of the Beamline I05.
	
	\item[Author Contributions] S.Z. conceived the research project. H.Z., J.L., F.W., W.C., X.C., J.A., P.D., M.D., A.L. and S.Z. performed the ARPES measurements and analyzed the ARPES data. K.L., S.W., H.Z., J.L. and G.C. prepared the samples. K.L., S.W., H.Z., J.L. and G.C. performed the AFM measurements. Y.W., Z.S. performed the calculations. T.S., P.Y. and W.D. contributed to the discussions. K.W. and T.T. prepared BN crystals. H.Z., J.L., and S.Z. wrote the manuscript, and all authors commented on the manuscript.
	
	\item[Competing Interests] The authors declare that they have no competing financial interests.
	
	\item[Correspondence and requests for materials] should be addressed to Guorui Chen (chenguorui@sjtu.edu.cn) and Shuyun Zhou (email: syzhou@mail.tsinghua.edu.cn).
	
\end{addendum}

\end{document}